%% file: WL-comb.tex
\documentclass[11pt]{article}
\usepackage[utf8]{inputenc}
\usepackage[left=1in,right=1in]{geometry}
\usepackage{authblk}
\usepackage[inline]{showlabels}
\usepackage[pdfstartview=FitH,pdfpagemode=None]{hyperref}
\usepackage{amsmath}
\usepackage{amsfonts}
\usepackage{cite}
\title{Combinatorics of Wilson loops in $\mathcal{N}=4$ SYM theory}
\author[1,2]{Wolfgang M\"uck \thanks{E-mail: \texttt{mueck@na.infn.it}}}
\affil[1]{Dipartimento di Fisica ``Ettore Pancini", Universit\`a degli Studi di Napoli ``Federico II" \authorcr Via Cintia, 80126 Napoli, Italy}
\affil[2]{Istituto Nazionale di Fisica Nucleare, Sezione di Napoli \authorcr Via Cintia, 80126 Napoli, Italy}
\date{}
\input{Definitions}

\begin{document}
\maketitle
\begin{abstract}
The theory of Wilson loops for gauge theories with unitary gauge groups is formulated in the language of symmetric functions. The main objects in this theory are two generating functions, which are related to each other by the involution that exchanges an irreducible representation with its conjugate. Both of them contain all information about the Wilson loops in arbitrary representations as well as the correlators of multiply-wound Wilson loops. This general framework is combined with the results of the Gaussian matrix model, which calculates the expectation values of $\frac12$-BPS circular Wilson loops in $\mathcal{N}=4$ Super-Yang-Mills theory. General, explicit, formulas for the connected correlators of multiply-wound Wilson loops in terms of the traces of symmetrized matrix products are obtained, as well as their inverses. It is shown that the generating functions for Wilson loops in mutually conjugate representations are related by a duality relation whenever they can be calculated by a Hermitian matrix model.
\end{abstract}
%%%%%%%%%%%%%%%%%%%%%%%%%%%%%%%%%%%%%%%%%%%%
\input{Intro}

%%%%%%%%%%%%%%%%%%%%%%%%%%%%%%%%%%%%%%%%%%%%
\input{WL}

%%%%%%%%%%%%%%%%%%%%%%%%%%%%%%%%%%%%%%%%%%%%
\input{Concs}

%%%%%%%%%%%%%%%%%%%%%%%%%%%%%%%%%%%%%%%%%%%%
%
\section*{Acknowledgements}
I would like to thank Anthonny Canazas Garay and Alberto Faraggi for their collaboration in an earlier project.
This work was supported in part by  the INFN, research initiative STEFI.
%%%%%%%%%%%%%%%%%%%%%%%%%%%%%%%%%%%%%%%%%%%%
%%%%%%%%%%%%%%%%%%%%%%%%%%%%%%%%%%%%%%%%%%%%
%\bibliographystyle{JHEPnotes}
%\bibliography{MM}
\bibliography{WL-comb}
\end{document}

%% file: Definitions.tex
\numberwithin{equation}{section}

\newcommand{\ie}{i.e.,\ }
\newcommand{\eg}{e.g.,\ }

\newcommand{\Tr}{\operatorname{Tr}}

\newcommand{\e}[1]{\operatorname{e}^{#1}}

\newcommand{\vev}[1]{\left\langle #1 \right\rangle}

%% file: Intro.tex
\section{Introduction}\label{sec: intro}

Wilson loops contain a lot of information about the dynamics of gauge theories and are important probes of non-perturbative physics. 
For example, in non-Abelian gauge theories, they can serve as an order parameter for confinement. Therefore, one is interested in theoretical tools and methods that allow to calculate Wilson loops exactly or as asymptotic $1/N$-expansions, beyond the planar approximation \cite{tHooft:1973alw, Brezin:1977sv, Itzykson:1979fi}. In the past two decades, holographic dualities \cite{Maldacena:1998im, Rey:1998ik, Drukker:1999zq}, integrability \cite{Minahan:2002ve} and localization techniques \cite{Pestun:2007rz, Pestun:2016zxk} have provided a wealth of new solutions, in particular for highly symmetric loop configurations in supersymmetric gauge theories.
A paradigmatic case is the case of $\frac12$-BPS circular Wilson loops in $\mathcal{N}=4$ Super-Yang-Mills theory with gauge group U$(N)$ or SU$(N)$. On the one hand, the holographic dual fully captures the planar approximation in the limit of large 't~Hooft coupling $\lambda$ \cite{Drukker:2005kx, Yamaguchi:2006te, Yamaguchi:2006tq, Gomis:2006sb, Lunin:2006xr, Gomis:2006im}, and a lot of effort has been dedicated to obtain corrections in $1/\lambda$ \cite{Forste:1999qn, Drukker:2000ep, Semenoff:2001xp, Kruczenski:2008zk, Faraggi:2011bb, Faraggi:2011ge, Faraggi:2014tna, Faraggi:2016ekd, Horikoshi:2016hds, Forini:2017whz, Aguilera-Damia:2018bam, Aguilera-Damia:2018twq, Medina-Rincon:2019bcc, David:2019lhr}. On the other hand, localization reduces the calculation of the Wilson loops to the solution of a Gaussian matrix model \cite{Erickson:2000af, Drukker:2000rr, Akemann:2001st, Hartnoll:2006is, Fiol:2013hna}, which, in principle, is exact in both, $\lambda$ and $N$ and provides an easier path to an asymptotic $1/N$-expansion \cite{Okuyama:2006jc, Chen-Lin:2016kkk, Gordon:2017dvy, Okuyama:2017feo, CanazasGaray:2018cpk, Okuyama:2018aij, Fiol:2018yuc}. Wilson loops in $\mathcal{N}=4$ Super-Yang-Mills theory with fewer symmetries have been studied, \eg in \cite{Zarembo:2002an, Drukker:2006ga, Drukker:2007dw, Drukker:2007qr, Forini:2015bgo}.

Localization can be applied more generally in $\mathcal{N}=2$ Super-Yang-Mills theories. Interested readers are referred to the recent paper \cite{Billo:2019fbi} and references therein. 

The purpose of this paper is to further develop a recent result \cite{CanazasGaray:2019mgq}, which relates the connected correlators of multiply-wound $\frac12$-BPS Wilson loops in $\mathcal{N}=4$ Super-Yang-Mills theory \cite{Okuyama:2018aij} to the exact solution of the corresponding Hermitian matrix model. This relation was worked out explicitly in \cite{Okuyama:2018aij} for the connected $n$-loop correlators with $n\leq 4$ and was generalized in \cite{CanazasGaray:2019mgq} to any $n$ by recognizing the combinatorical pattern. In \cite{CanazasGaray:2019mgq}, it was conjectured that the relation has a deeper, group-theoretical, origin. It will be shown here that this is indeed true. To achieve this goal, it turns out to be most natural and effective to employ the framework of symmetric functions. Symmetric functions have already been used in the matrix model solution of the $\frac12$-BPS Wilson loops in general representations \cite{Fiol:2013hna}. However, the generality of this framework does not seem to be widely appreciated. In fact, it allows to define the generating function(s) for Wilson loops in general representations, which will be done in this paper following the nice account of \cite{Marino:2005sj}. The framework of symmetric functions can also be translated to the languages of bosons or fermions in two dimensions \cite{Marino:2005sj}, which have their analogues in the context of matrix models.

The rest of the paper is organized as follows. In Section~\ref{WLgen}, the generating functions are defined, which contain all information on the expectation values of Wilson loops in arbitrary representations as well as the correlators of multiply-wound Wilson loops. Moreover, the connected correlators are defined. The generating functions for the connected correlators satisfy an interesting involution property, which is discussed in Section~\ref{involution}. 
The general framework is related, in Section~\ref{MM}, to the results of the Gaussian matrix model that evaluates, by localization, the $\frac12$-BPS Wilson loops of $\mathcal{N}=4$ SYM theory. 
This will result in explicit formulas for the connected correlators of multiply-wound Wilson loops in terms of the traces of symmetrized matrix products, and \emph{vice versa}. Finally, Section~\ref{Concs} contains some concluding comments.

Before starting, let us briefly introduce some notation. A partition $\lambda \vdash n$ is a weakly increasing (or weakly decreasing) set of positive integers $\lambda_i$ ($i=1,2,\ldots$) such that $\sum_i \lambda_i = |\lambda|= n$. The cardinality of $\lambda$ is denoted by $l(\lambda)$. Often, the notation $\lambda =\prod_i i^{a_i}$ is used, meaning that $\lambda$ contains the integer $i$ $a_i$ times. A partition $\lambda$ specifies the cycle type of a permutation and thus defines a conjugacy class $C_\lambda$ of the permutation group $\mathfrak{S}_n$. Defining the centralizer size by 
\begin{equation}
\label{Intro:z}
	z_\lambda = \prod\limits_i (a_i!\, i^{a_i})~,
\end{equation}
we have that $|C_\lambda| = |\lambda|!/z_\lambda$ is the size of the conjugacy class, \ie the number of permutations of cycle type $\lambda$.

Symmetric functions are crucial in this paper. Readers not familiar with them should consult a standard reference such as \cite{Macdonald:1995} or the lecture notes \cite{Lascoux}. Let $e_n$, $h_n$, and $p_n$ be the elementary, complete homogeneous and power-sum polynomials of degree $n$, respectively. For $a \in \{e,h,p\}$, given a partition $\lambda$, we define
$a_\lambda = \prod_i a_{\lambda_i}$. These functions form bases of symmetric functions (on some countably infinite alphabet). There are three additional classical bases, the basis of monomials, $m_\lambda$, the Schur basis, $s_\lambda$, and the ``forgotten'' basis, $f_\lambda$. Their role is captured best by considering the Hall inner product, $\vev{\cdot,\cdot}$, or the Cauchy kernel. The monomial basis is the adjoint of the complete homogeneous basis, $\vev{m_\lambda,h_\nu}=\delta_{\lambda\nu}$, the forgotten basis is the adjoint of the elementary basis, $\vev{f_\lambda,e_\nu}=\delta_{\lambda\nu}$, whereas the power-sum basis and the Schur basis satisfy $\vev{p_\lambda,p_\nu}=z_\lambda\delta_{\lambda\nu}$ and $\vev{s_\lambda,s_\nu}=\delta_{\lambda\nu}$, respectively. The Schur functions are related to the monomials by the Kostka matrix \cite{Macdonald:1995}.\footnote{The Kostka matrix was used in \cite{Fiol:2013hna} to obtain the Wilson loops in irreducible representations (Schur basis) from the matrix model solution (monomial basis), but we will not use it here.}

%% file: WL.tex
\section{Wilson loop generating functions}
\label{WLgen}

This section will follow the account of \cite{Marino:2005sj}. 
Consider a gauge theory with gauge group U$(N)$ or SU$(N)$. We are interested in the expectation value of a Wilson loop in an arbitrary irreducible representation of the gauge group, expressed in terms of symmetric polynomials. The irreducible representations are uniquely labelled by partitions $\lambda$, and their characters are given by the Schur polynomials. 

To start, let $U$ be the holonomy of the gauge connection for a single Wilson loop, an ``open'' Wilson loop, so to say. To use the language of symmetric polynomials, take $U$ diagonal, $U=\operatorname{diag}(u_1,u_2,\ldots)$ and denote by $u= (u_1,u_2,\ldots)$ the alphabet of its eigenvalues.\footnote{We will formally consider a countably infinite set of diagonal entries, almost all of which are zero.} 
It is obvious that the $n$-fold multiply-wound Wilson loop
\begin{equation}
\label{WL:Un.pn}
	\Tr U^n = \sum_i u_i^n = p_n(u)
\end{equation}
is given by the power-sum symmetric polynomial of degree $n$ in the eigenvalues.  Furthermore, we introduce an alphabet of real numbers $y = (y_1,y_2,\ldots)$ and define the two generating functions\footnote{$H(y)$ is the Cauchy kernel. It is also known as the Ooguri-Vafa operator \cite{Ooguri:1999bv}. $E(y)$ is the image of $H(y)$ under the involution that exchanges the elementary and the complete functions, as will become evident in \eqref{WL:EH2}.}
\begin{equation}
\label{WL:EH}
	E(y) = \prod\limits_{i,j} (1+y_i u_j)~, \qquad H(y) = \prod\limits_{i,j} \frac1{1-y_i u_j}~. 
\end{equation}
Expanding these as formal power series in $y$ and $u$ yields
\begin{equation}
\label{WL:EH2}
	E(y) = \sum\limits_{\lambda} e_\lambda(y) m_\lambda(u)~,\qquad  H(y) = \sum\limits_{\lambda} h_\lambda(y) m_\lambda(u)~.
\end{equation}
The sums are over all partitions $\lambda$. Obviously, the generating functions satisfy $E(y)H(-y)=1$.\footnote{This involution property was first mentioned in the context of Wilson loops in \cite{Hartnoll:2006is} for the special case of the generating functions of the totally symmetric and totally anti-symmetric representations.}

By the Cauchy identity \cite{Macdonald:1995}, we can express the generating functions in the Schur basis, 
\begin{equation}
\label{WL:Cauchy}
	E(y) = \sum\limits_\lambda  s_\lambda(y) s_{\lambda'}(u)~, \qquad  H(y) = \sum\limits_\lambda s_\lambda(y) s_\lambda(u)~,
\end{equation}
where
\begin{equation}
\label{WL:Tr.U}
	s_\lambda(u) = \Tr_\lambda(U) 
\end{equation}
is, by definition, the Wilson loop in the irreducible representation $\lambda$. Moreover, $\lambda'$ is the representation conjugate to $\lambda$, obtained by taking the transpose Young diagram. 

Starting from \eqref{WL:EH}, a short calculation shows that 
\begin{equation}
\label{WL:log.H}
	\ln H(y) = \sum_{n=1}^\infty \frac1{n} p_n(y) p_n(u)~.
\end{equation}
Then, exponentiating \eqref{WL:log.H} yields
\begin{equation}
\label{WL:H.p}
	H(y) = \sum\limits_\lambda \frac{1}{z_\lambda} p_\lambda(y) p_\lambda(u)~, 
\end{equation}
where we recognize in the power-sum functions $p_\lambda(u)$ the products of multiply-wound Wilson loops
\begin{equation}
\label{WL:prod.loops}
	p_\lambda(u) = \prod\limits_{i=1}^{l(\lambda)} p_{\lambda_i}(u) = \prod\limits_{i=1}^{l(\lambda)} \Tr U^{\lambda_i}~.
\end{equation}
Similarly, from $\ln E(y) = - \ln H(-y)$ we get
\begin{equation}
\label{WL:E.p}
	E(y) = \sum\limits_\lambda \frac{(-1)^{l(\lambda)}}{z_\lambda} p_\lambda(-y) p_{\lambda}(u)~. 
\end{equation}

Taking expectation values of the equations above, we obtain the following relations
\begin{subequations}
\begin{align}
\label{WL:ZH1}
	\qquad Z(y) = \vev{H(y)} &= \sum\limits_\lambda h_\lambda(y) \vev{m_\lambda(u)} \\
	&= \sum\limits_\lambda s_\lambda(y) \vev{s_\lambda(u)} \\
\label{WL:ZH3}
	&= \sum\limits_\lambda \frac{1}{z_\lambda} p_\lambda(y) \vev{p_\lambda(u)}~,
\end{align}
\end{subequations}
and
\begin{subequations}
\begin{align}
\label{WL:ZE1}
	\qquad Z'(y) = \vev{E(y)} &= \sum\limits_\lambda e_\lambda(y) \vev{m_\lambda(u)} \\
	&= \sum\limits_\lambda s_\lambda(y) \vev{s_{\lambda'}(u)} \\
\label{WL:ZE3}
	&= \sum\limits_\lambda \frac{(-1)^{l(\lambda)}}{z_\lambda} p_\lambda(-y) \vev{p_\lambda(u)}~.
\end{align}
\end{subequations}
Therefore, $Z(y)$ and $Z'(y)$ are generating functions for the expectation values of Wilson loops in any irreducible representation, $\vev{s_\lambda(u)}$, for the correlators of multiply-wound Wilson loops, $\vev{p_\lambda(u)}$, and also for the Wilson loop expectation value in monomial representations, $\vev{m_\lambda(u)}$. We note that $Z(y)$ and $Z'(y)$ are completely equivalent.  Expanded in the Schur basis, they generate the Wilson loops in the irreducible representations conjugate to each other. 
Another observation is that they contain all information on the Wilson loops, because the Schur functions (or the power-sum funtions or the monomial functions) form a complete basis of symmetric functions. For example, the correlator of two Wilson loops in irreducible representations $\lambda$ and $\nu$ can be decomposed according to the Littlewood-Richardson rule \cite{Macdonald:1995}
\begin{equation}
\label{WL:corr.irred}
	\vev{s_\lambda(u) s_\nu(u)} = \sum\limits_\mu c^\mu_{\lambda,\nu} \vev{s_\mu(u)}~.
\end{equation}

Finally, we define the connected correlators of multiply-wound Wilson loops by taking the logarithms of $Z$ and $Z'$\footnote{One may similarly define the ``connected'' expectation values in the monomial and the Schur bases, but their meaning is rather formal.}
\begin{align}
\label{WL:WH} 
	W(y) &= \ln Z(y) = \sum\limits_\lambda \frac{1}{z_\lambda} p_\lambda(y) 
		\vev{p_\lambda(u)}_\text{conn}~,\\
\label{WL:WE} 
	W'(y) &= \ln Z'(y) = \sum\limits_\lambda \frac{(-1)^{l(\lambda)}}{z_\lambda} p_\lambda(-y) 
		\vev{p_\lambda(u)}_\text{conn}~.
\end{align}
Taking $y=(z,0,0,\ldots)$, $W(y)$ and $W'(y)$ reduce to the generating functions of the totally symmetric and totally anti-symmetric Wilson loops, respectively.

\section{Involution property}
\label{involution}

A straightforward application of the symmetric-function formulation allows to generalize the recent observation \cite{Okuyama:2018aij, CanazasGaray:2019mgq} that the generating functions for $\frac12$-BPS Wilson loops in the totally symmetric and totally antisymmetric representations in $\mathcal{N}=4$ SYM theory are related to each other by an involution that changes the signs of the generating parameter ($y$ in our case) and of $N$. 
This involution property was proved recently \cite{Fiol:2018yuc} to be a general property for Wilson loops in representations conjugate to each other. Our proof presented below, which applies to the generating functions $W(y)$ and $W'(y)$, is essentially equivalent to the proof in \cite{Fiol:2018yuc}, section 3, which considers directly the (unnormalized) Wilson loops.  

In our notation, the involution property mentioned above reads\footnote{The functions $J_S$ and $J_A$ of \cite{Okuyama:2018aij, CanazasGaray:2019mgq} are given by $\frac1N W(y)$ and $\frac1N W'(y)$, respectively, with $y=(z,0,0,\ldots)$.}
\begin{equation}
\label{C:dual}
	W\left(y;\frac1N\right) = W'\left(-y;-\frac1N\right)~,
\end{equation} 
where we have added $1/N$ as a parameter. Clearly, \eqref{C:dual} is reminiscent of the involution property $E(y)H(-y)=1$ mentioned below \eqref{WL:EH2}, but that property is not sufficient to establish \eqref{C:dual}. The reason for this is simply that taking the expectation value does not commute with the logarithm in $\ln H(y) = - \ln E(-y)$. Comparing \eqref{WL:WH} with \eqref{WL:WE}, we also need
\begin{equation}
\label{C:dual.p}
	\vev{p_\lambda(u)}_{\text{conn};\frac1N} = (-1)^{l(\lambda)} \vev{p_\lambda(u)}_{\text{conn};-\frac1N}~,
\end{equation}
which is, a priori, far from obvious. However, \eqref{C:dual.p} is certainly true in those cases, in which the calculation of the Wilson loops can be mapped to a general, interacting, Hermitian one-matrix model \cite{Ambjorn:1992gw}. In these cases, the connected correlators of multiply-wound Wilson loops have a genus expansion of the form \cite{Ambjorn:1992gw, Okuyama:2018aij}
\begin{equation}
\label{C:genus}
	\vev{p_\lambda(u)}_{\text{conn};\frac1N} = N^{2-l(\lambda)} \sum\limits_{g=0}^\infty N^{-2g} C_{g,l(\lambda)} (\lambda)~,
\end{equation}
where the genus-$g$ contributions $C_{g,l(\lambda)}$ are independent of $N$. Therefore, the involution property \eqref{C:dual} holds in the case of the $\frac12$-BPS Wilson loops in $\mathcal{N}=4$ SYM theory discussed in this paper, and, more generally, in the $\mathcal{N}=2$ theories considered in \cite{Billo:2019fbi}.

\section{BPS Wilson loops in $\mathcal{N}=4$ SYM theory}
\label{MM}

Localization maps the calculation of BPS Wilson loops, in the case of $\mathcal{N}=2$ theories, to the solution of a matrix model \cite{Pestun:2007rz}. The case of $\frac12$-BPS circular Wilson loops in $\mathcal{N}=4$ SYM is particularly simple, because the matrix model is Gaussian. Considering $Z'(y)$, we have to calculate
\begin{equation}
\label{WL:Z.prime.mm}
	Z'(y) = \vev{\prod\limits_i \det \left( 1 + y_i \e{X}\right)}_{\mathrm{mm}}~.
\end{equation} 
We refer the reader to \cite{Fiol:2013hna,CanazasGaray:2018cpk,Okuyama:2018aij} for details of the matrix model and the calculation. The solution, in the case of the gauge group U$(N)$, is 
\begin{equation}
\label{WL:Z.prime.sol}
	Z'(y) = \det \left[ \sum\limits_{n=0}^\infty e_n(y) A_n \right]~,
\end{equation} 
with an $N \times N$ matrix $A_n$, the expression of which can be found in \cite{Okuyama:2018aij,CanazasGaray:2019mgq}. In what follows, we will not need $A_n$ explicitly, but we shall derive general formulas that relate the connected correlators of multiply-wound Wilson loops to the traces of symmetrized products of the matrices $A_n$.

First, take the logarithm of \eqref{WL:Z.prime.sol}, which yields after some calculation\footnote{To manipulate the multiple sums that appear after taking the logarithm, one can use the tools described in \cite{CanazasGaray:2019mgq}.} 
\begin{equation}
\label{WL:W.prime.sol}
	W'(y) = \sum\limits_\lambda e_\lambda(y)  \frac{(-1)^{l(\lambda)-1}}{z_\lambda} [l(\lambda)-1]!
	\left( \prod\limits_i \lambda_i\right) \Tr\left[ A_{(\lambda_1} A_{\lambda_2} \cdots A_{\lambda_{l(\lambda)})} \right]~.  
\end{equation}
From here, there are several ways to proceed. A formal path is to equate \eqref{WL:W.prime.sol} with \eqref{WL:WE} and use the Hall inner product in order to project both sides to the desired basis. Using the power-sum basis, this yields
\begin{equation}
\label{WL:W.p}
	\vev{p_\mu(u)}_\text{conn} = (-1)^{|\mu|+l(\mu)} \sum\limits_\lambda \frac{(-1)^{l(\lambda)-1}}{z_\lambda} 
	\left(\prod_i \lambda_i\right) [l(\lambda)-1]! \vev{e_\lambda, p_\mu} \Tr\left[ A_{(\lambda_1} A_{\lambda_2} \cdots A_{\lambda_{l(\lambda)})} \right]~.
\end{equation} 
Similarly, using the forgotten basis, $f_\lambda$, we obtain
\begin{equation}
\label{WL:W.f}
	\Tr\left[ A_{(\lambda_1} A_{\lambda_2} \cdots A_{\lambda_{l(\lambda)})} \right] = 
		\frac{(-1)^{l(\lambda)-1}}{[l(\lambda)-1]!} \frac{z_\lambda}{\prod_i \lambda_i} \sum\limits_\mu 
		\frac{(-1)^{|\mu|+l(\mu)}}{z_\mu} \vev{p_\mu,f_\lambda} \vev{p_\mu(u)}_\text{conn}~.
\end{equation} 
Equations \eqref{WL:W.p} and \eqref{WL:W.f} are fairly easy to implement on a computer algebra system such as \textsc{SageMath} \cite{sagemath}, but they hide the important feature that the indices on their left hand sides can be considered as a set. One would expect that the structure of the expansion on the right hand sides should depend only of the cardinality of this set, but not on its entries.   
Therefore, it is better to proceed differently. Let us use the equality of $E(y)$ in \eqref{WL:EH2} and \eqref{WL:E.p} together with \eqref{WL:W.prime.sol} to establish that
\begin{equation}
\label{WL:M.conn}
	\Tr\left[ A_{(\lambda_1} A_{\lambda_2} \cdots A_{\lambda_{l(\lambda)})} \right] = \frac{(-1)^{l(\lambda)-1}}{[l(\lambda)-1]!} 
	\frac{z_\lambda}{\prod_i \lambda_i} \vev{m_\lambda(u)}_\text{conn}~.
\end{equation}
Here, the connected expectation value of the monomial is a formal object, but if we are able to express the monomials in terms of the power-sum functions, then we have achieved our goal. Using the notation $\lambda = \prod_i i^{a_i}$, we have 
\begin{equation}
\label{WL:z.id}
	\frac{z_\lambda}{\prod_i \lambda_i} m_\lambda = \prod_i a_i !\, m_\lambda = \Phi_\lambda~,
\end{equation} 
which is called an augmented monomial. It can be expressed in the power-sum basis as follows \cite{Lascoux}. For a partition $\lambda$ with $l(\lambda)=n$, let us identify $\Phi_{\vec{k}}$ with $\Phi_\lambda$, if the $n$-dimensional vector $\vec{k}$ is a permutation of $\lambda$. Next, let $\nu =\{ \nu_1, \ldots, \nu_r\}$ be a partition of the set $\{1,\ldots ,n\}$.\footnote{This means that $\nu_1 \cup \ldots \cup \nu_r = \{1,\ldots ,n\}$ and all the $\nu_i$ are disjoint. $\nu$ is also called a set-partition or a decomposition.} Furthermore, define the ($r$-dimensional) vector
\begin{equation}
\label{WL:decomp.vec}
	\vec{k}_\nu = \left( \sum\limits_{i \in \nu_1} k_i, \ldots , \sum\limits_{i \in \nu_r} k_i \right)~.
\end{equation}
Then, with $\mathcal{M}(\nu)$ denoting the M\"obius function on the lattice of set-partitions,
\begin{equation}
\label{WL:Mobius}
	\mathcal{M}(\nu) = \prod\limits_{i=1}^r (-1)^{l(\nu_i)-1} [l(\nu_i)-1]!~,
\end{equation} 
we have
\begin{equation}
\label{WL:mon.to.pow.sum}
	\Phi_{\vec{k}} = \sum_{\nu\in\mathcal{P}(n)} \mathcal{M}(\nu)\, p_{\vec{k}_\nu}~,
\end{equation}
where the sum is over the lattice of set partitions, $\mathcal{P}(n)$. Notice that $\vec{k}_\nu$ is generally not a partition, but we simply understand $p_\nu=\prod_i p_{\nu_i}$ for any set $\nu$. 

Using these facts, \eqref{WL:M.conn} becomes
\begin{equation}
\label{WL:M.conn.2}
	\Tr \left[ A_{(\vec{k})} \right] \equiv \Tr\left[ A_{(k_1} A_{k_2} \cdots A_{k_n)} \right] = \frac{(-1)^{n-1}}{(n-1)!} 
	\sum\limits_{\nu\in\mathcal{P}(n)} \mathcal{M}(\nu)\, \vev{p_{\vec{k}_\nu}(u)}_\text{conn}~.
\end{equation}
The important point here is to notice that the entries of $\vec{k}$ appear on the right hand side only in $p_{\vec{k}_\nu}$, whereas the expansion coefficients are independent of them, which is just the property described above. Therefore, in order to find the coefficients in the expansion in \eqref{WL:M.conn.2} for any $\vec{k}$, it is sufficient to evaluate \eqref{WL:M.conn} for $\vec{k}=(1^n)$. This yields  
\begin{equation}
\label{WL:A1.n}
	\Tr\left( A_1^n\right) = \frac{(-1)^{n-1}}{(n-1)!} n! \vev{m_{(1^n)}(u)}_\text{conn} = 
	(-1)^{n-1} n \vev{e_n(u)}_\text{conn}~.
\end{equation}
Now we can use the relation 
\begin{equation}
\label{WL:en.p}
	e_n = \sum\limits_{\lambda\vdash n} \frac{(-1)^{|\lambda|+l(\lambda)}}{z_\lambda} p_\lambda
\end{equation}
to rewrite \eqref{WL:A1.n} as 
\begin{equation}
\label{WL:A1.new}
	\Tr\left( A_1^n\right) = n \sum\limits_{\lambda\vdash n} \frac{(-1)^{l(\lambda)-1}}{z_\lambda} 
	\vev{p_\lambda(u)}_\text{conn}~.
\end{equation}
Thus, taking into account that the left hand side of \eqref{WL:M.conn.2} is a symmetric function of the $k_i$, we find
\begin{equation}
\label{WL:tr.sym.A}
	\Tr \left[ A_{(\vec{k})} \right] = n 
	 \sum\limits_{\lambda\vdash n} \frac{(-1)^{l(\lambda)-1}}{z_\lambda} \vev{\widetilde{p_{\vec{k}_\lambda}}}_\text{conn}~,
\end{equation} 
where the tilde denotes symmetrization of the $k$'s, 
\begin{equation}
\label{WL:k.sym}
	\widetilde{p_{\vec{k}_\lambda}} = \frac1{n!} \sum\limits_{\sigma \in \mathfrak{S}_n} p_{\sigma(\vec{k})_\lambda}~.
\end{equation}
Equation \eqref{WL:tr.sym.A} is precisely the result found in \cite{CanazasGaray:2019mgq}. 

Without proof, I state here the inverse relation of \eqref{WL:tr.sym.A}. Let $\lambda =\prod_i i^{a_i}$ and let $\mathcal{P}_\lambda$ be the set of those set-partitions of $\{1,\ldots,n\}$ that contain $a_i$ subsets of size $i$. The cardinality of $\mathcal{P}_\lambda$, \ie the number of set partitions of $\{1,\ldots,n\}$ with $a_i$ subsets of size $i$, is given by \cite{NIST}
\begin{equation}
\label{WL:num.set.part}
	|\mathcal{P}_\lambda| = \frac{|\lambda|!}{\prod_i[ (i!)^{a_i} a_i!]}~.
\end{equation}
With this information, the inverse of \eqref{WL:tr.sym.A} is given by
\begin{equation}
\label{WL:pk}
	\vev{p_{\vec{k}}(u)}_\text{conn} = \sum\limits_{\lambda \vdash n} (-1)^{l(\lambda)-1} [l(\lambda)-1]!\, |\mathcal{P}_\lambda| 
	\Tr  \left [ \widetilde{A_{(\vec{k}_\lambda)}} \right]~,
\end{equation}
where $n=l(\vec{k})$ and the tilde denotes the symmetrization of the elements of $\vec{k}$, as above. 

I have checked with \textsc{SageMath} \cite{sagemath} that \eqref{WL:pk} and \eqref{WL:tr.sym.A} agree with \eqref{WL:W.p} and \eqref{WL:W.f}, respectively, for values up to $|\vec{k}|=8$. For higher values, the evaluation of \eqref{WL:pk} and \eqref{WL:tr.sym.A} requires some time because of the permutations involved.

%% file: Concs.tex
\section{Conclusions}
\label{Concs}

In this paper, the theory of Wilson loops for gauge theories with unitary gauge groups has been formulated in the language of symmetric functions. The main objects in this theory are the generating functions $Z(y)$ and $Z'(y)$, which are related to each other by the involution that exchanges an irreducible representation with its conjugate. The logarithms of $Z(y)$ and $Z'(y)$ define the connected Wilson loop correlators. Each of these generating functions contains all information about the Wilson loops in any irreducible representation of the gauge group, as well as on the correlators of multiply-wound Wilson loops. 
If the connected correlators of multiply-wound Wilson loops possess a genus expansion, which is true when the Wilson loop expectation values can be calculated by a Hermitian matrix model, then the involution property \eqref{C:dual} holds, for a simultaneous change of the signs of $y$ and $N$. 
Furthermore, we have applied this general theory to the results of the matrix model that calculates the expectation values of $\frac12$-BPS circular Wilson loops in $\mathcal{N}=4$ Super-Yang-Mills theory and obtained explicit and general formulas for the connected  correlators of multiply-wound Wilson loops in terms of the traces of symmetrized matrix products, and \emph{vice versa}, generalizing the results of \cite{Okuyama:2018aij, CanazasGaray:2019mgq}. It would be interesting to apply the framework of symmetric functions to 
the Hermitian matrix models that appear through localization in $\mathcal{N}=2$ Super-Yang-Mills theories. It would also be worthwhile to explore the implications of the symmetric-function formulation of Wilson loops for the gauge groups O$(N)$ and Sp$(N)$.